\documentstyle{article}

\author{L.S.F. Olavo\\
Departamento de Fisica ,Universidade de Brasilia - UnB\\
70910-900, Brasilia - D.F. - Brazil}
\title{Quantum Mechanics as a Classical Theory\\
VIII: Second Quantization}

\begin{document}

\maketitle
\begin{abstract}
We continue in this paper our program of rederiving all quantum mechanical
formalism from the classical one. We now turn our attention to the
derivation of the second quantized equations, both for integral and
half-integral spins. We then show that all the quantum results may be
derived using our approach and also show the interpretation suggested by
this derivation. This paper may be considered as a first approach to the
study of the quantum field theory beginning by the same classical ideas we
are supporting since the first paper of this series.
\end{abstract}

\section{Introduction}

In this paper we are interested in showing that the concept of second
quantization might be built within our purely classical reconstruction of
quantum mechanics. More than that, when deriving the relevant results, we
are led to a much clearer picture of the `creation' and `annihilation'
operators action.

We will develop our study in the realm of two important systems where second
quantization methods may be applied---which are fermionic and bosonic
systems. We show that all the mathematical results of ordinary quantum
mechanics may be retrieved by our methods.

We begin the next section obtaining the quantum Schr\"odinger equation for
the harmonic oscillator problem using the {\it classical} Hamilton's
equations together with Liouville's equation. This will be attained with the
use of the Wigner-Moyal Infinitesimal Transformation\cite{eu1,eu2,eu3}. Then
we perform a canonical transformation in the classical phase space to get a
new Hamiltonian and, using again the Wigner-Moyal Infinitesimal
Transformation, we derive the canonically transformed Schr\"odinger
equation. We then show that this equation is precisely the `second
quantized' one. This result shows that, with our method, it is possible to
quantize physical systems in any generalized phase-space---we left, however,
the development of this result to a future paper.

The third section is devoted to the study of classical second quantization
of fermionic systems. The same procedures used for bosons will be applied
for fermions and we will retrieve again the formalism encountered in the
literature.

In the last section we make our final conclusions.

\section{Harmonic Oscillator}

We begin with the {\it classical} harmonic oscillator defined by the
hamiltonian
\begin{equation}
\label{1}H=\frac 1{2m}p^2+\frac{m\omega ^2}2q^2
\end{equation}
and the {\it classical} statistical Liouville equation
\begin{equation}
\label{2}\frac{dF}{dt}=\frac{\partial F}{\partial t}+\frac{dq}{dt}\frac{%
\partial F}{\partial q}+\frac{dp}{dt}\frac{\partial F}{\partial p}=0.
\end{equation}
Using Hamilton's equations
\begin{equation}
\label{3}\frac{dq}{dt}=\frac{\partial H}{\partial p}=\frac pm\ ;\ \frac{dp}{%
dt}=-\frac{\partial H}{\partial q}=-m\omega ^2q\
\end{equation}
we rewrite (\ref{2}) as
\begin{equation}
\label{4}\frac{\partial F}{\partial t}+\frac pm\frac{\partial F}{\partial q}%
-m\omega ^2q\ \frac{\partial F}{\partial p}=0.
\end{equation}

We now define the density function as the Infinitesimal Wigner-Moyal
transformation
\begin{equation}
\label{5}\rho \left( q-\frac{\delta q}2,q+\frac{\delta q}2;t\right) =\int
F\left( q,p;t\right) e^{\frac i\hbar p\delta q}dp
\end{equation}
and apply it upon equation (\ref{4}) to get
\begin{equation}
\label{6}-i\hbar \frac{\partial \rho }{\partial t}-\frac{\hbar ^2}m\frac{%
\partial ^2\rho }{\partial q\partial (\delta q)}+m\omega ^2q\delta q\rho =0
\end{equation}
We now suppose that this density function might be written as
\begin{equation}
\label{7}\rho \left( q-\frac{\delta q}2,q+\frac{\delta q}2;t\right) =\Phi
^{\dagger }\left( q-\frac{\delta q}2;t\right) \Phi \left( q+\frac{\delta q}%
2;t\right)
\end{equation}
and use the fact that the function $\Phi (q;t)$ is complex and might be
written as
\begin{equation}
\label{8}\Phi \left( q;t\right) =R\left( q;t\right) e^{\frac i\hbar S\left(
q;t\right) }
\end{equation}
where $R$ and $S$ are real functions.

It is then possible to take expression (\ref{7})---with (\ref{8}) for the
functions $\Phi $---into expression (\ref{6}) and, retaining only terms in
the zeroth and first order on the infinitesimal parameter $\delta q$, to
derive the following equation
$$
\left[ \frac{\partial (R^2)}{\partial t}+\frac \partial {\partial q}\left(
R^2\frac{\partial S/\partial q}m\right) \right] +
$$
\begin{equation}
\label{9}+R^2\frac{i\delta q}\hbar \frac \partial {\partial q}\left[ \frac{%
\partial S}{\partial t}+\frac 1{2m}\left( \frac{\partial S}{\partial q}%
\right) ^2-\frac{\hbar ^2}{2mR}\frac{\partial ^2R}{\partial q^2}+\frac{%
m\omega ^2}2q^2\right] =0.
\end{equation}
The real and imaginary terms have to be each identically zero and we get the
two equations
\begin{equation}
\label{10}\frac{\partial (R^2)}{\partial t}+\frac \partial {\partial
q}\left( R^2\frac{\partial S/\partial q}m\right) =0
\end{equation}
and
\begin{equation}
\label{11}\frac{\partial S}{\partial t}+\frac 1{2m}\left( \frac{\partial S}{%
\partial q}\right) ^2-\frac{\hbar ^2}{2mR}\frac{\partial ^2R}{\partial q^2}+
\frac{m\omega ^2}2q^2=0
\end{equation}
where in this last equation we made the arbitrary constant equal to zero
like in our other papers\cite{eu1}.

We may show by straightforward calculations that these equations are
precisely those we obtain when we substitute expression (\ref{8}) in the
Schr\"odinger equation
\begin{equation}
\label{12}-\frac{\hbar ^2}{2m}\frac{\partial ^2\Phi }{\partial q^2}+\frac
12m\omega ^2q^2\Phi =i\hbar \frac{\partial \Phi }{\partial t}
\end{equation}
where now we immediately identify expression (\ref{10}) as the continuity
equation. We then conclude that these two sets of equations (\ref{10},\ref
{11}) and (\ref{12}) have the same mathematical content and so use them
interchangeably. Therefore, we have shown that it is possible to derive the
harmonic oscillator quantum equation from classical mechanics by means of
the Infinitesimal Wigner-Moyal Transformation.

We now make a canonical transformation in the {\it classical} phase space
defined by
\begin{equation}
\label{13}q_1=\left[ \frac p{\sqrt{2\hbar m\omega }}+i\sqrt{\frac{m\omega }{%
2\hbar }}q\right] \ ;\ p_1=\left[ \frac p{\sqrt{2\hbar m\omega }}-i\sqrt{
\frac{m\omega }{2\hbar }}q\right]
\end{equation}
to get the new hamiltonian
\begin{equation}
\label{14}H_1=\left( \hbar \omega \right) q_1p_1
\end{equation}
and the new Liouville equation ($F$ is now a function of $q_1$ and $p_1$)
\begin{equation}
\label{15}\frac{\partial F}{\partial t}+\hbar \omega q_1\frac{\partial F}{%
\partial q_1}-\hbar \omega p_1\ \frac{\partial F}{\partial p_1}=0,
\end{equation}
where we used Hamilton's equations
\begin{equation}
\label{16}\frac{dq_1}{dt}=\hbar \omega q_1\ ;\ \frac{dp_1}{dt}=-\hbar \omega
p_1.
\end{equation}

With the definition
\begin{equation}
\label{17}\rho \left( q_1-\frac{\delta q_1}2,q_1+\frac{\delta q_1}2;t\right)
=\int F\left( q_1,p_1;t\right) e^{\frac i\hbar p_1\delta q_1}dp_1
\end{equation}
for the density function we follow the same steps as above to get the
equation
\begin{equation}
\label{18}\frac{\partial \rho }{\partial t}+\hbar \omega q_1\frac{\partial
\rho }{\partial q_1}-\hbar \omega \frac \partial {\partial (\delta
q_1)}\left( \delta q_1\rho \right) =0.
\end{equation}
Imposing the format
\begin{equation}
\label{19}\rho \left( q_1-\frac{\delta q_1}2,q_1+\frac{\delta q_1}2;t\right)
=\Phi ^{\dagger }\left( q_1-\frac{\delta q_1}2;t\right) \Phi \left( q_1+
\frac{\delta q_1}2;t\right)
\end{equation}
upon the density function and using the expression
\begin{equation}
\label{20}\Phi \left( q_1;t\right) =R\left( q_1;t\right) e^{\frac i\hbar
S\left( q_1;t\right) }
\end{equation}
we get, with the same considerations as above, the following pair of
equations
\begin{equation}
\label{21}\frac{\partial (R^2)}{\partial t}+\hbar \omega q_1\frac{\partial
(R^2)}{\partial q_1}+\hbar \omega R^2=0
\end{equation}
and
\begin{equation}
\label{22}\frac \partial {\partial q_1}\left[ \frac{\partial S}{\partial t}%
+\hbar \omega q_1\frac{\partial S}{\partial q_1}\right] =0.
\end{equation}

The pair of equations (\ref{21},\ref{22}) is equivalent---in the sense
defined above---to the Schr\"odinger equation
\begin{equation}
\label{23}\hbar \omega \left[ -i\hbar q_1\frac{\partial \Phi }{\partial q_1}+
\frac{i\hbar }2\Phi \right] =i\hbar \frac{\partial \Phi }{\partial t}
\end{equation}
which can be rewritten as
\begin{equation}
\label{24}\hbar \omega \left( a^{\dagger }a+\frac{[a,a^{\dagger }]}2\right)
\Phi =i\hbar \frac{\partial \Phi }{\partial t},
\end{equation}
if we make the identification
\begin{equation}
\label{25}a^{\dagger }=q_1\ ;\ a=-i\hbar \frac \partial {\partial q_1}.
\end{equation}

Equation (\ref{24}) is nothing but the second quantized equation for the
harmonic oscillator when we make the canonical transformation (\ref{13}) in
the operator space and apply it to the Schr\"odinger equation (\ref{12}).
These calculations show that we may perform the canonical transformation in
the classical (functions) or quantum (operators) phase space as we wish,
since the results will be exactly the same.

Equation (\ref{23}) is a differential equation and we might solve it to
obtain the amplitude $\Phi $ as
\begin{equation}
\label{26}\Phi (q_1;t)=q_1^ne^{-iEt/\hbar }\Phi _0,
\end{equation}
where $\Phi _0$ is a constant usually called the vacuum state. The energy
may also be obtained and we get
\begin{equation}
\label{27}E=\hbar \omega \left( n+\frac 12\right)
\end{equation}
as expected.

In the language of operators $a$ and $a^{\dagger }$, the amplitude in (\ref
{26}) may be written as
\begin{equation}
\label{28}\left| \Phi (q_1)\right\rangle =\left( a^{\dagger }\right)
^ne^{-iEt/\hbar }\left| \Phi _0\right\rangle ,
\end{equation}
which can be found in the literature\cite{Dirac}.

We may go one step further which is very instructive for the sake of
interpretation. Based on the expressions (\ref{13}) for the variable $q_1$
we may define
\begin{equation}
\label{29}\cos \theta =\frac p{\sqrt{2\hbar m\omega }}\ ;\ \sin \theta =
\sqrt{\frac{m\omega }{2\hbar }}q
\end{equation}
which implies
\begin{equation}
\label{30}\tan \theta =\frac{m\omega q}p
\end{equation}
that defines the angle $\theta $ as the phase difference between the
movements on the $q$ and $p$-axes, in the sense that the classical solutions
imply
\begin{equation}
\label{30a}p=\sqrt{2mE}\cos (\omega t+\theta )\ ;\ q=\sqrt{\frac{2E}{m\omega
^2}}\sin (\omega t+\theta ),
\end{equation}
and so%
$$
\tan \theta =\frac{m\omega q(t_0)}{p(t_0)}.
$$
This gives us the value
\begin{equation}
\label{31}q_1=e^{i\theta }
\end{equation}
and (\ref{26}) may be written as
\begin{equation}
\label{32}\Phi _n(q_1,t)=e^{in\theta }e^{-iEt/\hbar }\Phi _0.
\end{equation}

In terms of the operators $a$ and $a^{\dagger }$ we get
\begin{equation}
\label{33}a^{\dagger }\left| \Phi _n(q_1)\right\rangle =e^{i(n+1)\theta
}e^{-iEt/\hbar }\left| \Phi _0\right\rangle \ ;\ a\left| \Phi
_n(q_1)\right\rangle =e^{i(n-1)\theta }e^{-iEt/\hbar }\left| \Phi
_0\right\rangle
\end{equation}
apart from normalizations. These equations fix the interpretation of the
operators $a$ and $a^{\dagger }$. These operators act as excitation or
deexcitation of the normal modes defined upon phase space for the harmonic
oscillator. Transforming to these normal modes signify that we go from our
phase space ellipse defined by the hamiltonian (\ref{1}) into an hyperbole
defined by the hamiltonian (\ref{14}).

\section{Half-Integral Spin Particles}

We now pass to the study of particles with half-integral spin. We will base
this study on the equations previously obtained by ourselves\cite{eu6,eu7}
for such particles.

The functions involved in this study were%
$$
S_1=\frac 1{2\omega }\left( \frac 1mp_xp_y+m\omega ^2xy\right) \ ;\
S_2=\frac 1{4\omega }\left[ m\omega ^2\left( x^2-y^2\right) +\frac 1m\left(
p_x^2-p_y^2\right) \right]
$$

\begin{equation}
\label{34}S_3=\frac 12\left( xp_y-yp_x\right) \ ;\ S_0=\frac 1{2\omega
}\left[ \frac 1m\left( p_x^2+p_y^2\right) +m\omega ^2\left( x^2+y^2\right)
\right]
\end{equation}
and
\begin{equation}
\label{35}S^{\prime 2}=\frac 14S_0^2,
\end{equation}
where we have put
\begin{equation}
\label{36}\sqrt{\frac \alpha \beta }=m\omega .
\end{equation}

The problem consists in making diagonal the operator related with function $%
S^{\prime 2}$ and the operator related with one of the functions $S_i$. To
make the operator $S^{\prime 2}$ diagonal is the same as to make diagonal
the operator $\widehat{S}_0$ because of (\ref{35}). Indeed, we have shown%
\cite{eu7} that if
\begin{equation}
\label{37}\widehat{S}_0\psi =\hbar \lambda \psi ,
\end{equation}
then
\begin{equation}
\label{38}\widehat{S}^2\psi =\hbar ^2\left( \frac{\lambda -1}2\right) \left(
\frac{\lambda +1}2\right) \psi ,
\end{equation}
where we have used
\begin{equation}
\label{38.a}\widehat{S}^2=\widehat{S}^{\prime 2}-\hbar ^2/4
\end{equation}
or, if we put
\begin{equation}
\label{39}\widehat{S}^2\psi =\hbar ^2\left( \frac N2\right) \left( \frac
N2+1\right) \psi
\end{equation}
then
\begin{equation}
\label{40}N=\lambda -1.
\end{equation}

We now introduce the following canonical transformation
\begin{equation}
\label{41}q_1=\left[ \frac{p_x}{\sqrt{2\hbar m\omega }}+i\sqrt{\frac{m\omega
}{2\hbar }}x\right] \ ;\ p_1=\left[ \frac{p_x}{\sqrt{2\hbar m\omega }}-i
\sqrt{\frac{m\omega }{2\hbar }}x\right]
\end{equation}
and
\begin{equation}
\label{42}q_2=\left[ \frac{p_y}{\sqrt{2\hbar m\omega }}+i\sqrt{\frac{m\omega
}{2\hbar }}y\right] \ ;\ p_2=\left[ \frac{p_y}{\sqrt{2\hbar m\omega }}-i
\sqrt{\frac{m\omega }{2\hbar }}y\right]
\end{equation}
to write
\begin{equation}
\label{43}S_0^{\prime }=\hbar \left( q_1p_1+q_2p_2\right)
\end{equation}
where the prime indicates that the functions is written in the transformed
coordinate system.

The functions $S_i$ are, after the canonical transformation,
\begin{equation}
\label{44}S_1^{\prime }=\frac \hbar {2i}\left[ \left( q_1^2-q_2^2\right)
+\left( p_1^2-p_2^2\right) \right] \ ;\ S_2^{\prime }=\frac \hbar 2\left(
q_1p_1-q_2p_2\right)
\end{equation}
and
\begin{equation}
\label{45}S_3^{\prime }=\frac \hbar {2i}\left( q_1p_2-q_2p_1\right) .
\end{equation}
We then choose to make diagonal both $S_0^{\prime }$ and $S_2^{\prime }$
\begin{equation}
\label{46}S_2^{\prime }=\frac \hbar 2\left( q_1p_1-q_2p_2\right) \ ;\
S_0^{\prime }=\hbar \left( q_1p_1+q_2p_2\right) .
\end{equation}

When making the quantization defined by the application of the Infinitesimal
Wigner-Moyal Transformation (as above) we find the two operators
\begin{equation}
\label{47}\widehat{S}_2^{\prime }=\frac \hbar 2\left( a_1^{\dagger
}a_1-a_2^{\dagger }a_2\right) \ ;\ \widehat{S}_0^{\prime }=\hbar \left(
a_1^{\dagger }a_1+a_2^{\dagger }a_2+1\right)
\end{equation}
with the conventional definitions for the second quantization operators
\begin{equation}
\label{48}a_1^{\dagger }=q_1\ ;\ a_1=-i\hbar \frac \partial {\partial q_1}\
;\ a_2^{\dagger }=q_2\ ;\ a_2=-i\hbar \frac \partial {\partial q_2}.
\end{equation}

In spite of working with the operator $\widehat{S}_0^{\prime }$ we may, as
in expression (\ref{38.a}), work with the operator
\begin{equation}
\label{49}\widehat{N}=\hbar \left( a_1^{\dagger }a_1+a_2^{\dagger
}a_2\right)
\end{equation}
with eigenvalue
\begin{equation}
\label{50}N=\lambda -1.
\end{equation}

In terms of the total `angular momentum' we have the similar problem to make
diagonal the operators
\begin{equation}
\label{51}\widehat{S}_2^{\prime }=\frac \hbar 2\left( a_1^{\dagger
}a_1-a_2^{\dagger }a_2\right) \ ;\ \widehat{S}^{\prime 2}=\hbar \left( \frac{
\widehat{N}}2\right) \left( \frac{\widehat{N}}2+1\right) .
\end{equation}

The related eigenvectors are given by the tensor product
\begin{equation}
\label{52}\left| \Phi _{n_1}\Phi _{n_2}\right\rangle =(a_1^{\dagger
})^{n1}(a_2^{\dagger })^{n2}e^{-iEt/\hbar }\left| \Phi _0^1\Phi
_0^2\right\rangle
\end{equation}
which are precisely the (not normalized) eigenvectors found in the
literature \cite{Baym,Schwinger}. The same considerations about the relative
phase in phase space may be approached with the same methods which yield the
same interpretation for the operators $a_i$ and $a_i^{\dagger },$ $i=1,2$.

\section{Conclusions}

We have thus shown how to go from the formalism of classical mechanics to
the second quantization one---both for integral and half-integral spin
particles.

As was already said, this paper serves for a double intention. The first one
is completeness. We have taken the task of showing that {\it all} quantum
mechanical formalism may be derived from the classical one by means of the
infinitesimal Wigner-Moyal Transformation. This was done.

The second intention is much more ambitious. As everyone knows, second
quantization is the touchstone of quantum field theory. We hope that the
study of second quantization within the interpretation here proposed may
help one to understand the infinities occurring in the formalism of quantum
field theory and so gives us the means to develop a more acceptable theory.
This second aim will be dealt with in a future paper.


\begin{thebibliography}{9}
\bibitem{eu1}  Olavo, L. S. F., quant-ph/9503020

\bibitem{eu2}  Olavo, L. S. F., quant-ph/9503021

\bibitem{eu3}  Olavo, L. S. F., quant-ph/9503022

\bibitem{Dirac}  Dirac, P. A. M., ''{\it Directions in Physics}'' (John
Wiley and Sons, New York, 1978).

\bibitem{eu6}  Olavo, L. S. F., quant-ph/9503024

\bibitem{eu7}  Olavo, L. S. F., quant-ph/9503025

\bibitem{Baym}  Baym, G. ''{\it Lectures on Quantum Mechanics}'' (The
Benjamin/Cummings Publishing Co., CA, 1973).

\bibitem{Schwinger}  Schwinger, J., AEC report NYO-3071 (1952).
\end{thebibliography}
\end{document}